\documentclass[11pt]{article}

\usepackage[top=1in, bottom=1in, left=1in, right=1in]{geometry}

\usepackage{amsmath}
\usepackage{amsfonts}
\usepackage{wasysym}
\usepackage{graphicx}
\usepackage{comment}

\usepackage{hyperref}

\numberwithin{equation}{section}

\def\tn{\textnormal}

\begin{document}



\thispagestyle{empty}

\hfill 0903.0123 [hep-ph]

\hfill February 28, 2009

\addvspace{45pt}

\begin{center}

\Large{\textbf{Retrofitting Models of Inflation}}
\\[35pt]
\large{Ben Kain}
\\[10pt]
\textit{Department of Physics and Department of Mathematics and Computer Science }
\\ \textit{Santa Clara University}
\\ \textit{Santa Clara, CA 95053, USA}
\\[10pt] 
bkain@scu.edu
\end{center}

\addvspace{35pt}

\begin{abstract}
\noindent I use the method of retrofitting, developed by Dine, Feng and Silverstein, to generate the scale of inflation dynamically, allowing it to be naturally small.  This is a general procedure that may be performed on existing models of supersymmetric inflation.  I illustrate this idea on two such models, one an example of $F$-term inflation and the other an example of $D$-term inflation.
\end{abstract} 

\newpage


\section{Introduction}
\label{intro}

The results of WMAP \cite{wmap5} have put inflation on firm ground.  However, inflation model building still faces many challenges.  One in particular is the scale of inflation.  To understand this, let me review how the inflation scale is determined in slow roll inflation \cite{inf}.  During inflation, the potential energy is roughly constant,
\begin{equation}
	V=V_\tn{inf},
\end{equation}
so $V_\tn{inf}^{1/4}$ is the inflation scale.  As the inflaton rolls down its potential, $V$, it produces a curvature perturbation with a spectrum given by
\begin{equation} \label{cp}
	{\cal P}_\zeta = \frac{1}{24\pi^2M_p^4} \frac{V}{\epsilon},
\end{equation}
where $M_p=2.4\times10^{18}$ GeV is the reduced Planck mass and $\epsilon$ is one of the two standard slow roll parameters, 
\begin{equation} \label{srp}
	\epsilon = M_p^2\frac{1}{2} \left(\frac{V'}{V}\right)^2, \qquad \eta=M_p^2\frac{V''}{V},
\end{equation}
with a prime denoting differentiation with respect to the inflaton.  In (\ref{cp}) the right hand side is to be evaluated at the scale of horizon exit, which for most models of inflation corresponds to 50--60 $e$-folds before the end of inflation \cite{dhll}.  The number of $e$-folds, $N(\phi_*)$, from an inflaton value $\phi=\phi_*$ to the end of inflation at $\phi=\phi_e$ is given by
\begin{equation} \label{nefolds}
	N(\phi)=-\frac{1}{M^2_p}\int_{\phi_*}^{\phi_e} \frac{V}{V'} d\phi.
\end{equation}
WMAP has made a measurement of the spectrum of curvature perturbations \cite{wmap5},
\begin{equation}
	{\cal P}_\zeta^{1/2} = 4.86\times10^{-5},
\end{equation}
which when combined with (\ref{cp}) gives the CMB constraint,\footnote{This is also commonly referred to as the COBE or WMAP normalization.}
\begin{equation} \label{cmbcon}
	V^{1/4} = (0.027M_p)\epsilon^{1/4} = (6.6\times10^{16} \tn{ GeV})\epsilon^{1/4}.
\end{equation}
In practice, ones uses (\ref{nefolds}) to determine the inflaton value for 50--60 $e$-folds before the end of inflation and then plugs this value into (\ref{cmbcon}).  In other words, (\ref{cmbcon}) is to be evaluated during inflation, when $V=V_\tn{inf}$.  Since a necessary condition for inflation is $\epsilon \ll 1$, which requires $\epsilon^{1/4} < 1$, we are lead to the bound
\begin{equation} \label{vinfbound}
	V_\tn{inf}^{1/4} < 0.027M_p =  6.6\times10^{16} \tn{ GeV}.
\end{equation}
Inflation models are often built such that $V_\tn{inf}$ is made up of dimensionful parameters placed into the scalar potential by hand.  In this paper I will take the fundamental scale to be the (reduced) Planck scale.  Under this assumption, once these parameters are fit to the CMB constraint they can take on unnaturally small values.\footnote{Of course, if I were to take the fundamental scale to be the GUT scale then this is not necessarily an issue.}

I will use the method of retrofitting, developed by Dine, Feng and Silverstein \cite{dfs}, to generate the small parameters dynamically.  This is a general procedure that may be applied to previously constructed models of inflation.  It works by introducing a new supersymmetric sector, for example a pure $SU(n)$ supersymmetric gauge theory, that at a scale $\Lambda$ becomes strongly coupled.  By coupling specific fields to this new supersymmetric sector and integrating out the massive gaugino condensates below the scale of condensation, a nonperturbative contribution to the superpotential is induced, introducing the naturally small scale $\Lambda$.  I will demonstrate how to retrofit models of inflation with two specific examples, the first, in section \ref{rfterm}, is an example of $F$-term inflation, and the second, in section \ref{rdterm}, is an example of $D$-term inflation.  Both of these models have an inflation scale around $10^{15}$ GeV, and so are models that have parameters only a few orders of magnitude below the Planck scale.  Regardless, they are simple, representative models of supersymmetric inflation that act as good examples for demonstrating the retrofitting technique.  In both cases I will start by reviewing the models with the small parameters put in by hand and then I will retrofit them. 


\section{Retrofitting a Model of \texorpdfstring{$F$}{F}-term Inflation}
\label{rfterm}


\subsection{The Model}
\label{rfsusy}

The model of $F$-term inflation I will retrofit was first constructed in \cite{cllsw, dss} and is an example of hybrid inflation.  The superpotential is
\begin{equation} \label{fw}
	W=\Phi\left(\lambda X\overline{X} - \mu^2 \right),
\end{equation}
where $\Phi$ is a singlet superfield whose scalar component is the inflaton, $X$ and $\overline{X}$ are conjugate superfields oppositely charged under all symmetries so that $X\overline{X}$ is invariant and $\lambda$ and $\mu$ are constants.  Since there is only one dimensionful parameter in the model, $\mu$, we should expect it to set the scale of inflation.  We will see shortly that this is the case.

The scalar potential is
\begin{equation} \label{fsp}
	V=|\lambda\phi|^2\left( |x|^2 + |\bar{x}|^2 \right) + \left|\lambda x\bar{x} - \mu^2\right|^2,
\end{equation}
where I have used lowercase symbols for the scalar components of the superfields.  If $X$ and $\overline{X}$ are charged under a gauge symmetry there will also be a $D$-term.  However, since we will be considering the $D$-flat direction $|x|=|\bar{x}|$, it does not need to be considered.  The second term in (\ref{fsp}) may be written 
\begin{equation}
	\bigl| \lambda x\bar{x} - \mu^2 \bigr|^2 = \bigl| |\lambda x\bar{x}|  - e^{i\theta}|\mu|^2 \bigr|^2,  
\end{equation}
from which it is easy to see that $\langle\theta\rangle=0$, to which it will henceforth be set.  Without loss of generality I will take $\lambda$ and $\mu$ to be real and positive.  The scalar potential can now be written as
\begin{equation}
	V=\lambda^2|\phi|^2\left( |x|^2 + |\bar{x}|^2 \right) + \left(\lambda|x\bar{x}| - \mu^2\right)^2.
\end{equation}
The supersymmetric vacuum is at
\begin{equation}
	\langle\phi\rangle = 0, \qquad \langle |x| \rangle = \langle |\bar{x}| \rangle = \frac{\mu}{\sqrt{\lambda}}.
\end{equation}

When the inflaton is above its critical value, $|\phi| > |\phi_c| =\mu/\sqrt{\lambda}$, then $x=\bar{x}=0$ is a minimum and the potential in this minimum is
\begin{equation}
	V=\mu^4,
\end{equation}
which causes inflation.  In the notation of the introduction, $V_\tn{inf}^{1/4}=\mu$ is the scale of inflation.  A potential for the inflaton may be introduced by including the one-loop correction \cite{dss}, given by the well known formula
\begin{equation} \label{cw}
	V_\tn{1-loop} = \frac{1}{64\pi^2}\tn{STr}\left[{\cal M}^4\ln\left(\frac{{\cal M}^2}{Q^2}\right)\right] = \frac{1}{64\pi^2}\sum_{i,j} (-1)^{2j}(2j+1) m_{j,i}^4\ln\left(\frac{m_{j,i}^2}{Q^2}\right),
\end{equation}
where $m_{j,i}$ is the $i$-th eigenvalue from the mass matrix for particles of spin $j$ and $Q$ is a momentum cutoff.  The scalar fields have squared eigenmasses
\begin{equation}
	m^2_{\pm} = \lambda^2|\phi|^2 \pm \lambda\mu^2
\end{equation}
and the superpartner fermion squared eigenmasses are all equal to $\lambda^2|\phi|^2$.  All other contributions vanish during inflation since $x=\bar{x}=0$, leading to the one-loop scalar potential \cite{dss}
\begin{equation}
\begin{aligned}
	V = \mu^4 + \frac{\lambda^2}{32\pi^2} \biggl[2\mu^4\ln\left(\frac{\lambda^2|\phi|^2}{Q^2}\right) &+ \left(\lambda|\phi|^2+\mu^2\right)^2\ln\left(1+\frac{\mu^2}{\lambda|\phi|^2}\right) \\ &+\left(\lambda|\phi|^2-\mu^2\right)^2\ln\left(1-\frac{\mu^2}{\lambda|\phi|^2}\right)\biggr].
\end{aligned}	
\end{equation}
If during inflation we assume $|\phi| \gg |\phi_c|=\mu/\sqrt{\lambda}$ then the potential reduces to
\begin{equation} \label{ftermpot}
	V = \mu^4 \left[ 1 + \frac{\lambda^2}{16\pi^2}\ln\left(\frac{\lambda^2|\phi|^2}{Q^2}\right)\right].
\end{equation}

Now that we have an inflaton potential, following the outline in the introduction we can use the CMB constraint (\ref{cmbcon}) to determine the value of $\mu$.  This is done for potentials of the form (\ref{ftermpot}) in the appendix.  Borrowing the result (\ref{cmbpot}) and taking $\lambda=1$, I find
\begin{equation}
	\mu \sim 10^{-3} M_p \sim 10^{15}\tn{ GeV}.
\end{equation}


\subsection{Retrofitting}
\label{rf}

I will use the method of retrofitting \cite{dfs} to generate the small parameter in (\ref{fsp}), $\mu$, dynamically.\footnote{This model of inflation has been modified to include dynamically generated terms in \cite{ddr}, but in that paper they used a different technique.}  This method begins by introducing a new supersymmetric sector.  I will take this sector to be a pure $SU(n)$ supersymmetric gauge theory with superfield strength $W_\alpha$.  The superpotential (\ref{fw}) is then written without the $\mu$ term,
\begin{equation} \label{fw0}
	W_0=\lambda \Phi X \overline{X},
\end{equation}
and the new supersymmetric sector is included in the Lagrangian through the terms \cite{dfs, dm}
\begin{equation}
	\int d^2\theta\frac{\Phi}{4M_p} W^\alpha W_\alpha + \tn{h.c.}
\end{equation}
Once this supersymmetric sector becomes strongly coupled and confines it induces a nonperturbative contribution to the superpotential,
\begin{equation} \label{fwnp}
	W_\tn{np}=\Lambda^3 e^{-8\pi\Phi/b_0M_p} \sim \Lambda^3 -\frac{\Lambda^3}{M_p}\Phi + O(M_p^{-2}),
\end{equation}
where $b_0$ is the $\beta$-function for the $SU(n)$ gauge theory.  The complete superpotential is the sum of (\ref{fw0}) and (\ref{fwnp}) and is given by
\begin{equation}
	W=W_0 + W_\tn{np} \sim \Lambda^3 + \Phi\left(\lambda X \overline{X} -\frac{\Lambda^3}{M_p}\right).
\end{equation}
The constant term, $\Lambda^3$, does not play a role in global supersymmetry since it does not enter the scalar potential.  Comparing this superpotential with (\ref{fw}) we find
\begin{equation}
	\mu^2 \sim \frac{\Lambda^3}{M_p},
\end{equation}
and thus the dimensionful parameter that was placed in by hand in section \ref{rfsusy} has been dynamically generated and can be naturally small.


\section{Retrofitting a Model of \texorpdfstring{$D$}{D}-term Inflation}
\label{rdterm}

The retrofitted model of $D$-term inflation will turn out to be rather unsatisfying.  One of the primary motivations for $D$-term inflation is that the vacuum energy causing inflation comes entirely from the $D$-term of the scalar potential, and in so doing solves the supergravity $\eta$-problem.  Retrofitting the model requires introducing a new dimensionful parameter and effectively trading the vacuum energy coming from the $D$-term for vacuum energy coming from the $F$-term, defeating the original motivation.  This immediately introduces the $\eta$-problem.  As I will explain, the $\eta$-problem can be solved for certain choices of the K\"ahler potential, and is no longer solved for a general K\"ahler potential.


\subsection{The Model}
\label{rdsusy}

The model of $D$-term inflation that I will retrofit is the original model \cite{bdh} and, like the $F$-term model above, is an example of hybrid inflation.  The superpotential is
\begin{equation} \label{dw}
	W=\lambda \Phi X_+X_-,
\end{equation}
where $\Phi$ is a singlet superfield whose scalar component is the inflaton, $X_+$ and $X_-$ are charged under a $U(1)$ gauge symmetry with charges $+1$ and $-1$ and $\lambda$ is a constant.  The scalar potential is
\begin{equation} \label{dsp}
	V=|\lambda \phi^2|\left(|x_+|^2 +|x_-|^2\right) + \left| \lambda x_+ x_- \right|^2 +\frac{1}{2}g^2\left( |x_+|^2 - |x_-|^2 + \xi \right)^2,
\end{equation}
where the final term is the $D$-term, a Fayet-Iliopoulos term, $\xi$, has been included and $g$ is the $U(1)$ coupling constant.  Since the Fayet-Iliopoulos term is the only dimensionful parameter in the theory, we will find that it sets the scale of inflation.  Without loss of generality I will take $\lambda$ and $\xi$ to be real and positive.  The supersymmetric vacuum is at\footnote{If I take $\xi$ to be negative then the VEVs for $x_+$ and $x_-$ switch.}
\begin{equation}
	\langle \phi \rangle =  \langle x_+ \rangle = 0, \qquad \langle |x_-| \rangle = \sqrt{\xi}.
\end{equation}

When the inflaton is above its critical value, $|\phi| > |\phi_c| = g\sqrt{\xi}/\lambda$, then $x_+ = x_- = 0$ is a minimum and the potential in this minimum is
\begin{equation} \label{dinfscale}
	V=\frac{1}{2}g^2\xi^2,
\end{equation}
which causes inflation.  In the notation of the introduction, $V_\tn{inf}^{1/4}=\sqrt{g\xi}/2^{1/4}$ is the scale of inflation.  A potential for the inflaton may be introduced by including the one-loop correction, using (\ref{cw}).  The scalar fields have squared eigenmasses
\begin{equation}
	m^2_{\pm} = \lambda^2|\phi|^2 \pm g^2\xi
\end{equation}
and the superpartner fermion squared eigenmasses are all equal to $\lambda^2|\phi|^2$.  All other contributions vanish during inflation since $x_+=x_-=0$, leading to the one-loop scalar potential
\begin{equation}
\begin{aligned}
	V = \frac{1}{2}g^2\xi^2 + \frac{1}{32\pi^2} \biggl[ 2g^4\xi^2\ln\left(\frac{\lambda|\phi|^2}{Q^2}\right) 
	&+ \left(\lambda^2|\phi|^2+g^2\xi\right)^2\ln\left(1+\frac{g^2\xi}{\lambda^2|\phi|^2}\right) \\
	&+ \left(\lambda^2|\phi|^2-g^2\xi\right)^2\ln\left(1-\frac{g^2\xi}{\lambda^2|\phi|^2}\right) \biggr].
\end{aligned}	
\end{equation} 
If during inflation we assume $|\phi| \gg |\phi_c|=g\sqrt{\xi}/\lambda$ then the potential reduces to
\begin{equation}\label{dtermpot}
	V = \frac{1}{2}g^2\xi^2 \left[ 1 + \frac{g^2}{8\pi^2}\ln\left(\frac{\lambda|\phi|^2}{Q^2}\right)\right].
\end{equation}

Now that we have an inflaton potential, following the outline in the introduction we can use the CMB constraint (\ref{cmbcon}) to determine the value of $g\xi$.  This is done for potentials of the form (\ref{dtermpot}) in the appendix.  Borrowing the result (\ref{cmbpot}) and taking $g=1$, I find
\begin{equation}
	\sqrt{\xi} \sim 10^{-3} M_p \sim 10^{15}\tn{ GeV}.
\end{equation}
 

\subsection{Modifying the Model for Retrofitting}
\label{rd}

Before the above model can be retrofitted, it must be modified \cite{dfs}.  A new dimensionful parameter, $m$, will be introduced which will effectively replace the parameter $\xi$ in the inflation scale (though, as we will see, it will not amount to a complete replacement).  This new parameter will then be retrofitted.  

The modification begins by adding two new superfields, $A_+$ and $A_-$, with charges $+1$ and $-1$ under the $U(1)$ gauge symmetry.  These new superfields enter the superpotential through a mass term,\footnote{The negative sign in front of $m$ is immaterial, as can be seen from the scalar potential (\ref{modsp}).  The choice of a negative sign is for future convenience when retrofitting in section \ref{rd2}.}
\begin{equation} \label{dw2}
	W=\lambda \Phi X_+ X_- - mA_+ A_-.
\end{equation}
The scalar potential is
\begin{equation}
\begin{aligned} \label{modsp}
	V=&\lambda^2|\phi^2|\left(|x_+|^2 +|x_-|^2\right) + \lambda^2\left|x_+ x_- \right|^2 +m^2\left(|a_+|^2 + |a_-|^2\right) \\
	&+\frac{1}{2}g^2\left( |x_+|^2 - |x_-|^2 + |a_+|^2 - |a_-|^2 + \xi \right)^2,
\end{aligned}
\end{equation}
where, without loss of generality, I have taken $m$ to be real and positive.  The supersymmetric vacuum is at
\begin{equation}
	\langle \phi \rangle = \langle x_+ \rangle = \langle a_+ \rangle = \langle a_- \rangle = 0, \qquad \langle |x_-| \rangle = \sqrt{\xi}.
\end{equation}

For the inflaton above its critical value, $|\phi|>|\phi_c|=m/\lambda$, and for $\xi > m^2/g^2$, there is a minimum at 
\begin{equation} \label{drvevs}
	x_+ = x_- = a_+ = 0, \qquad |a_-|^2 = \xi - \frac{m^2}{g^2},
\end{equation}
and the potential in this minimum is
\begin{equation}
	V= m^2\left(\xi - \frac{m^2}{2g^2}\right).
\end{equation}
If we assume $\xi \gg m^2/2g^2$, then the scalar potential reduces to
\begin{equation} \label{dinfscale2}
	V=m^2\xi,
\end{equation}
which comes entirely from the $F$-term of the scalar potential.  The fact that the $F$-term of the scalar potential is contributing to the vacuum energy during inflation introduces the supergravity $\eta$-problem, which I will explain in the next subsection.

As before, to introduce an inflaton potential we include the one loop correction using (\ref{cw}).  From (\ref{drvevs}) we can see that not all of the charged fields have vanishing VEVs.  This allows for contributions from gauginos and the gauge boson.  However, since the VEVs in (\ref{drvevs}) are all $\phi$-independent, so too are the new contributions.  Ignoring these constant, $\phi$-independent contributions, the one loop potential is
\begin{equation}
\begin{aligned}
	V=m^2\xi + \frac{1}{32\pi^2}\biggr[ 2m^4\ln\left(\frac{\lambda^2|\phi|^2}{Q^2}\right) 
	&+ \left(\lambda^2|\phi|^2+m^2\right)^2\ln\left(1+\frac{m^2}{\lambda^2|\phi|^2}\right) \\
	 &+\left(\lambda^2|\phi|^2-m^2\right)^2\ln\left(1-\frac{m^2}{\lambda^2|\phi|^2}\right)
	\biggl].
\end{aligned}
\end{equation}
If during inflation we assume $|\phi| \gg |\phi_c|=m/\lambda$ then the potential reduces to\footnote{In the appendix I mention the problem of Planckian field values in these models.  This problem is remedied if the loop correction is suppressed (leading to a small value for $C$  in (\ref{phistar})).  Once $m$ is retrofitted it can be naturally small, which from (\ref{dmpot}) leads to a suppression of the loop correction.}
\begin{equation} \label{dmpot}
	V=m^2\xi\left[1+\frac{1}{16\pi^2}\frac{m^2}{\xi}\ln\left(\frac{\lambda^2|\phi|^2}{Q^2}\right)\right].
\end{equation}

If we take $\xi\sim O(M^2_p)$, we can see from (\ref{dinfscale2}) and (\ref{dmpot}) that the addition of the new fields has effectively replaced the inflation scale being determined by $g\xi$ (see (\ref{dinfscale})) with it being determined by $m$.  The retrofitting technique used in section \ref{rfterm} can now be used to generate $m$ dynamically, which will be described in section \ref{rd2}.


\subsection{Supergravity and the \texorpdfstring{$\eta$}{eta}-Problem}
\label{etaprob}

In the modified model the $F$-term of the scalar potential is contributing to the vacuum energy during inflation.  For this reason retrofitting this model is rather unsatisfying since one of the primary motivations for $D$-term inflation is that the vacuum energy comes entirely from the $D$-term.  To see why a contribution from the $F$-term is problematic we must consider supergravity corrections to the scalar potential.

For convenience, in this subsection I will set the reduced Planck mass equal to one: $M_p=1$.  In supergravity\footnote{When there is a fixed Fayet-Iliopoulos term, a nonstandard supergravity generalization of global supersymmetry should be used \cite{bdkv}.  However, I will use the standard supergravity generalization to explain the $\eta$-problem.} the $F$-term contribution to the scalar potential is given by
\begin{equation} \label{sugrasp}
	V_F = e^{K}\left[K^{m\bar{n}}\left(W_m + K_m W\right)\left(\overline{W}_{\bar{n}} + K_{\bar{n}} \overline{W} \right)-3|W|^2\right],
\end{equation}
where an unbarred subscript denotes differentiation with respect to a chiral superfield and a barred subscript denotes differentiation with respect to an antichiral superfield and $K^{m\bar{n}}$ is the inverse K\"ahler metric.  Plugging in the superpotential (\ref{dw2}) and a minimal K\"ahler potential,
\begin{equation}
	K=|\Phi|^2+|X_+|^2+|X_-|^2+|A_+|^2+|A_-|^2,
\end{equation}
and placing the fields at their VEVs (\ref{drvevs}), gives
\begin{equation} \label{eppot}
	V = e^{K} (m^2\xi) = e^\xi m^2\xi\left[1+|\phi|^2+O(|\phi|^4)\right],
\end{equation}
where I have assumed, as before, that $\xi \gg m^2/2g^2$.  This potential does not lead to inflation.  In particular, inflation requires $|\eta| \ll 1$, where $\eta$ is the slow roll parameter in (\ref{srp}), but from (\ref{eppot}) we find $|\eta| \sim 1$.  This is a generic consequence of the $F$-term contributing vacuum energy during inflation\footnote{The $F$-term model in section \ref{rfterm} solves the $\eta$-problem by a fortuitous cancellation that occurs for minimal K\"ahler potentials and a superpotential linear in the inflaton.} and is known as the $\eta$-problem.

One way to solve this problem is to use a non-minimal K\"ahler potential.  For example, a K\"ahler potential of the no-scale form that comes from string theory \cite{cllsw, stew, gmo},\footnote{For a more comprehensive discussion on inflation model building with such K\"ahler potentials in string derived models, see, for example, \cite{cllsw,glmk}.}
\begin{equation} \label{kp}
	K=-\ln\left(1-\sum\nolimits_a|\phi_a|^2\right).
\end{equation}
The $F$-term contribution to the scalar potential can be calculated using (\ref{sugrasp}) and is
\begin{equation}
	V_F = \sum\nolimits_a \left| W_a \right|^2 - 2\left( 1-\sum\nolimits_a|\phi_a|^2\right)^{-1}\left| W \right|^2 - \left|W-\sum\nolimits_a\phi_a W_a \right|^2.
\end{equation}
Plugging in the superpotential (\ref{dw2}) and placing the fields at their VEVs (\ref{drvevs}) we can immediately see that only the first term survives and reproduces the $F$-term from global supersymmetry, solving the $\eta$-problem.  

More generally, some fields could enter the K\"ahler potential minimally, so that we could have
\begin{equation} \label{kp2}
	K=-\ln\left(1-\sum\nolimits_a|\phi_a|^2\right) + \sum\nolimits_b|\phi_b|^2.
\end{equation}
For this to still solve the $\eta$-problem we we need at least the inflaton, $\phi$, and $a_+$ to enter the K\"ahler potential in the form (\ref{kp}), i.e. as $\phi_a$ fields \cite{stew}. I will take the rest of the fields to enter the K\"ahler potential minimally, i.e. as $\phi_b$ fields.  The $F$-term can again be calculated using (\ref{sugrasp}),  and plugging in the superpotential (\ref{dw2}) and placing the fields at their VEVs (\ref{drvevs}), I find
\begin{equation} \label{dsugravac}
	V_F = e^\xi m^2\xi,
\end{equation}
where, as before, I have assumed $\xi\gg m^2/2g$.  The exponential comes from $a_-$ entering the K\"ahler potential minimally .

The $D$-term contribution to the scalar potential is given by
\begin{equation} \label{dterm2}
\begin{aligned}
	V_D &= \frac{1}{2}g^2 \left(\sum\nolimits_m q_m K_m \phi_m + \xi \right)^2 \\
	&= \frac{1}{2}g^2\left[\sum\nolimits_a q_a  \left(1-\sum\nolimits_a|\phi_a|^2\right)^{-1}|\phi_a|^2 + \sum\nolimits_b q_b |\phi_b|^2 +\xi \right]^2,
\end{aligned}
\end{equation}
where the $q_m$ are the $U(1)$ charges and in the second line I plugged in the K\"ahler potential (\ref{kp2}).  Note that for the choice of the inflaton, $\phi$, and $a_+$ entering the K\"ahler potential as $\phi_a$ fields and the rest of the fields entering minimally as $\phi_b$ fields, when the fields are placed at their VEVs (\ref{drvevs}) the $D$-term (\ref{dterm2}) gives the same result as in global supersymmetry (remembering, of course, that the inflaton is uncharged).  Thus, when using (\ref{kp2}) as the K\"ahler potential, the $\eta$-problem is solved and the analysis of inflation at the supergravity level is the same as in global supersymmetry, along with the slight modification that the inflation scale is now given by (\ref{dsugravac}).

Finally, the K\"ahler potential (\ref{kp}) leads to a noncanonical kinetic term for the inflaton.  If we ignore the phase, a canonically normalized inflaton can be included in the scalar potential through the substitution
\begin{equation}
	|\phi| \rightarrow \tanh(|\phi|/\sqrt{2}) = \frac{1}{\sqrt{2}}|\phi| - \frac{1}{6\sqrt{2}}|\phi|^3 + O(|\phi|^5).
\end{equation}


\subsection{Retrofitting}
\label{rd2}

Retrofitting the parameter $m$ begins by introducing a new supersymmetric sector.  As before I will take this to be a pure $SU(n)$ gauge theory with superfield strength $W_\alpha$.  The superpotential (\ref{dw2}) is then written without the $m$ term,
\begin{equation} \label{dw0}
	W_0=\lambda X_+ X_-,
\end{equation}
and the new supersymmetric sector is included in the Lagrangian through the term \cite{dfs, dm}
\begin{equation}
	\int d^2\theta\frac{A_+A_-}{4M_p^2} W^\alpha W_\alpha + \tn{h.c.}
\end{equation}
Once the supersymmetric sector becomes strongly coupled and confines it induces a nonperturbative contribution to the superpotential,
\begin{equation} \label{dwnp}
	W_\tn{np}=\Lambda^3 e^{-8\pi A_+ A_- /b_0 M^2_p} \sim \Lambda^3 - \frac{\Lambda^3}{M_p^2}A_+ A_- + O(M_p^{-4}).
\end{equation}
The complete superpotential is the sum of (\ref{dw0}) and (\ref{dwnp}) and is given by
\begin{equation}
	W=W_0 + W_\tn{np} \sim \Lambda^3 + \lambda \Phi X_+ X_- - \frac{\Lambda^3}{M_p^2}A_+ A_-.
\end{equation}
The constant term, $\Lambda^3$, does not play a role in global supersymmetry since it does not enter the scalar potential.\footnote{In supergravity it can be tuned away.}  Comparing this superpotential with (\ref{dw2}) we find that
\begin{equation}
	m \sim \frac{\Lambda^3}{M_p^2}.
\end{equation}
From (\ref{dsugravac}) this means that the vacuum energy during inflation is now
\begin{equation}
	V\sim\left(\frac{\Lambda^3}{M_p^2}\right)^2 e^{\xi/M^2_p}\xi.
\end{equation}
For $\xi\sim O(M^2_p)$ the scale of inflation is effectively set by the dynamically generated scale $\Lambda$.  In this way the Fayet-Iliopoulos term, that was placed in by hand in section \ref{rdsusy}, has been effectively replaced by a dynamically generated one.


\section{Conclusion}
\label{con}

Dimensionful parameters in models of inflation are often put in by hand.  Some combination of these parameters set the scale of inflation and, when fit against the CMB constraint, can turn out to be unnaturally small.  I have used the method of retrofitting, developed by Dine, Feng and Silverstein \cite{dfs}, to generate these parameters dynamically, allowing them to be naturally small.  This is a general procedure that may be applied to existing models of inflation.  I have illustrated this idea with two such models, one an example of $F$-term inflation and the other an example of $D$-term inflation.  The retrofitted model of $D$-term inflation is rather unsatisfying since the vacuum energy during inflation no longer comes entirely from the $D$-term, introducing the $\eta$-problem, which can be solved for certain choices of the K\"ahler potential.  In the course of retrofitting the $D$-term model I have ignored the production of cosmic strings \cite{lrj}, which leads to various bounds on the parameters of the model.  It might be interesting to consider these bounds within the context of retrofitting.


\section*{Acknowledgments}
I cannot thank enough Linda M. Carpenter for her comments and for reading a draft of this paper.  I am also grateful to Rich Barber and Dennis Smolarski.


\appendix

\section{Scale of Inflation}
In this appendix I will determine the general form of the scale of inflation for the models considered above.  In both models, the scalar potential is of the form
\begin{equation} \label{gensp}
	V=V_\tn{inf}\left[1+\frac{C}{16\pi^2}\ln\left(\frac{\phi^2}{Q^2}\right)\right],
\end{equation}
where $C$ is a dimensionless constant and coefficients of $\phi$ have been absorbed into $Q$.  Inflation ends when either of the slow roll parameters in (\ref{srp}), $\epsilon$ and $|\eta|$, are equal to one or when $\phi$ reaches its critical value, whichever occurs first.  However, we do not need to know when this occurs becuase the integral (\ref{nefolds}) is dominated by $\phi_*$.  Evaluating (\ref{nefolds}) for the potential (\ref{gensp}) I find
\begin{equation} \label{phistar}
	\phi_* =(1.2)\sqrt{\left(\frac{C}{1}\right)\left(\frac{N}{55}\right)}M_p.
\end{equation}
Notice that $\phi_*$ is near the Planck scale for 50--60 $e$-folds .  This is problematic since these models are built at the level of an effective field theory where field values must be much lower than the Planck scale.  Using this value for the inflaton and applying the CMB constraint (\ref{cmbcon}) I find
\begin{equation} \label{cmbpot}
	V_\tn{inf}^{1/4}=(2.4\times10^{-3}M_p)\left[\left(\frac{C}{1}\right)\left(\frac{55}{N}\right)\right]^{1/4}=(5.7\times10^{15}\tn{ GeV})\left[\left(\frac{C}{1}\right)\left(\frac{55}{N}\right)\right]^{1/4}
\end{equation}
for the scale of inflation.



\begin{thebibliography}{99}
\bibitem{wmap5}
  E.~Komatsu {\it et al.}  [WMAP Collaboration],
  arXiv:0803.0547 [astro-ph].
\bibitem{inf}
	A.~R.~Liddle and D.~H.~Lyth,
  ``Cosmological inflation and large-scale structure,'' Cambridge, UK: Univ. Pr. (2000); 
  D.~H.~Lyth and A.~Riotto,
  Phys.\ Rept.\  {\bf 314}, 1 (1999)
  [arXiv:hep-ph/9807278].
\bibitem{dhll}
  S.~Dodelson and L.~Hui,
  Phys.\ Rev.\ Lett.\  {\bf 91}, 131301 (2003)
  [astro-ph/0305113];
  A.~R.~Liddle and S.~M.~Leach,
  Phys.\ Rev.\ D {\bf 68}, 103503 (2003)
  [astro-ph/0305263].
\bibitem{cllsw}
  E.~J.~Copeland, A.~R.~Liddle, D.~H.~Lyth, E.~D.~Stewart and D.~Wands,
  Phys.\ Rev.\  D {\bf 49}, 6410 (1994)
  [arXiv:astro-ph/9401011].
\bibitem{dss}
  G.~R.~Dvali, Q.~Shafi and R.~K.~Schaefer,
  Phys.\ Rev.\ Lett.\  {\bf 73}, 1886 (1994)
  [arXiv:hep-ph/9406319].
\bibitem{dfs}
  M.~Dine, J.~L.~Feng and E.~Silverstein,
  Phys.\ Rev.\  D {\bf 74}, 095012 (2006)
  [arXiv:hep-th/0608159].
\bibitem{ddr}
  S.~Dimopoulos, G.~R.~Dvali and R.~Rattazzi,
  Phys.\ Lett.\  B {\bf 410}, 119 (1997)
  [arXiv:hep-ph/9705348].
\bibitem{dm}
  M.~Dine and J.~D.~Mason,
  arXiv:0712.1355 [hep-ph].
\bibitem{bdh}
	P.~Binetruy and G.~R.~Dvali,
  Phys.\ Lett.\  B {\bf 388}, 241 (1996)
  [arXiv:hep-ph/9606342];
  E.~Halyo,
  Phys.\ Lett.\  B {\bf 387}, 43 (1996)
  [arXiv:hep-ph/9606423].
\bibitem{bdkv}
  P.~Binetruy, G.~Dvali, R.~Kallosh and A.~Van Proeyen,
  Class.\ Quant.\ Grav.\  {\bf 21}, 3137 (2004)
  [arXiv:hep-th/0402046].
\bibitem{stew}
  E.~D.~Stewart,
  Phys.\ Rev.\  D {\bf 51}, 6847 (1995)
  [arXiv:hep-ph/9405389];
\bibitem{gmo}
	M.~K.~Gaillard, H.~Murayama and K.~A.~Olive,
  Phys.\ Lett.\  B {\bf 355}, 71 (1995)
  [arXiv:hep-ph/9504307].
\bibitem{glmk}
  M.~K.~Gaillard, D.~H.~Lyth and H.~Murayama,
  Phys.\ Rev.\  D {\bf 58}, 123505 (1998)
  [arXiv:hep-th/9806157];
  B.~Kain,
  Nucl.\ Phys.\  B {\bf 800}, 270 (2008)
  [arXiv:hep-ph/0608279].
\bibitem{lrj}
  D.~H.~Lyth and A.~Riotto,
  Phys.\ Lett.\  B {\bf 412}, 28 (1997)
  [arXiv:hep-ph/9707273];
  R.~Jeannerot,
  Phys.\ Rev.\  D {\bf 56}, 6205 (1997)
  [arXiv:hep-ph/9706391].
\end{thebibliography}
\end{document}